\documentclass[10pt]{revtex4}
\usepackage{amsfonts}
\usepackage{amssymb,epsf}
\usepackage{latexsym}                                
\begin{document}

\title{Modified Friedmann Equations From Debye
Entropic Gravity}
\author{A.  Sheykhi$^{1,2}$ \footnote{
sheykhi@mail.uk.ac.ir} and Z. Teimoori$^{1}$}
\address{$^1$Department of Physics, Shahid Bahonar University, P.O. Box 76175, Kerman, Iran\\
         $^2$ Research Institute for Astronomy and Astrophysics of Maragha
         (RIAAM), P.O. Box 55134-441, Maragha, Iran}
\begin{abstract}
A remarkable new idea on the origin of gravity was recently proposed
by Verlinde who claimed that the laws of gravitation are no longer
fundamental, but rather emerge naturally as an entropic force. In
Verlinde derivation, the equipartition law of energy on the
holographic screen plays a crucial role. However, the equipartition
law of energy fails at the very low temperature. Therefore, the
formalism of the entropic force should be modified while the
temperature of the holographic screen is very low. Considering the
Debye entropic gravity and following the strategy of Verlinde, we
derive the modified Newton's law of gravitation and the
corresponding Friedmann equations which are valid in all range of
temperature. In the limit of strong gravitational field, i.e. high
temperature compared to Debye temperature, $T\gg T_D$, one recovers
the standard Newton's law and Friedmann equations. We also
generalize our study to the entropy corrected area law and derive
the dynamical cosmological equations for all range of temperature.
Some limits of the obtained results are also studied.

\end{abstract}

\maketitle
\section{Introduction}\label{Int}
Although gravity is the most universal force of nature, however, the
origin of it in quantum level is still unclear. This is due to the
fact that it is remarkably hard to combine gravity with quantum
mechanics compared with all the other forces, and hence the final
theory of the quantum gravity has not been established yet. The
universality of gravity suggests that its emergence should be
understood from general principles that are independent of the
specific details of the underlying microscopic theory.

According to Einstein's theory of general relativity, the concept of
gravity has strongly connected to the spacetime geometry. Indeed,
Einstein field equations tell us that the presence of energy or
stress causes the deformation of the spacetime geometry. In 1970's
Bekenstein and Hawking \cite{HB} discovered black holes
thermodynamics. With combination of quantum mechanics and general
relativity they predicted that a black hole behaves like a black
body, emitting thermal radiations, with a temperature proportional
to its surface gravity at the black hole horizon and with an entropy
proportional to its horizon area. The Hawking temperature and the
horizon entropy together with the black hole mass obey the first law
of thermodynamics, $dM=TdS$. Since the discovery of black hole
thermodynamics in 1970's, physicists have been speculating that
there should be a direct relation between thermodynamics and
Einstein equations. This is expected because the geometrical
quantities like horizon area and surface gravity are proportional to
entropy and horizon temperature, respectively. After Bekenstein and
Hawking a lot of works have been done to disclose the connection
between themodynamics and gravity \cite{B,D}. In 1995 Jacobson
\cite{Jac} put forward a great step and showed that the Einstein
field equation is just an equation of state for the spacetime and in
particular it can be derived from the proportionality of entropy and
the horizon area together with the fundamental relation $\delta
Q=TdS$. Following Jacobson, however, an overwhelming flood of papers
has appeared which attempt to show that there is indeed a deeper
connection between gravitational field equations and horizon
thermodynamics. It has been shown that, not only in Einstein gravity
but also in a wide variety of theories, the gravitational field
equations for the spacetime metric has a predisposition to
thermodynamic behavior. This result, first pointed out in
\cite{Pad}, has now been demonstrated in various theories including
f(R) gravity \cite{Elin} and cosmological setups
\cite{Cai2,Cai3,CaiKim,Wang,Cai33,Shey0,Shey1,Shey2}. For a recent
review on the thermodynamical aspects of gravity and complete list
of references see \cite{Pad0}.

Recently, Verlinde \cite{Ver} has invented a conceptual theory that
gravity is no longer fundamental, but is emergent. According to
Verlinde, one can start from the first principles, and gravity
appears as an entropic force naturally and unavoidably in a theory
in which space is emergent through a holographic scenario. Similar
discoveries are also made by Padmanabhan \cite{Padm} who observed
that the equipartition law for horizon degrees of freedom combined
with the Smarr formula leads to the Newton's law of gravity. In
addition, Verlinde's arguments reveal a fact that the key to
understanding gravity is information (or entropy). In Verlinde
derivation the holographic principle and the equipartition law of
energy play a crucial role. The holographic principle was originally
proposed by 't Hooft \cite{Hooft} and then developed in cosmology by
Susskind and others \cite{Sus}. According to this principle the
combination of quantum mechanics and gravity requires the three
dimensional world to be an image of data that can be stored on a two
dimensional projection much like a holographic image. The studies on
the entropic gravity scenario have arisen a lot attention recently
\cite{Cai4,Smolin,Li,Tian,Vancea,Modesto,Sheykhi1,BLi,Sheykhi2,Gu,other,mann}.

It is well-known that the equipartition law of energy for a system
of particles only valid for the situation where the kinetic energy
of the particles is much larger than the effective interacting
potential between particles. This means that the equipartition law
of energy break down at very low temperatures. It is found that
Debye model, which modified the equipartition law of energy, is in
good agreement with experimental results for most solid objects.
According to Verlinde, we know that the gravity can be explained as
an entropic force, it means that the gravity may have a statistical
thermodynamics explanation. Therefore, the formalism of the entropic
force should be modified while the temperature of the holographic
screen is very low. This means that Newtonian gravity takes a
different form in the background of an extreme weak gravitational
field. In the present work, inspired by the Debye's model in
statistical thermodynamics, we generalize the formalism of the
entropic gravity to the very law temperatures.

The outline of the present paper is as follows.  The modified
Newton's law of gravitation and the corresponding Friedmann
equations which are valid in all range of temperature are extracted
in the next section. Sec. III is devoted to the derivation of the
Entropy corrected Friedmann equations in Debye entropic force
scenario. The paper ends with a conclusion, which appears in Sec. IV.

\section{Debye Entropic Gravity and Friedmann Equation}\label{III}
Consider a closed holographic screen and a free particle of mass $m$
near it on the side that spacetime has already emerged. In Verlide's
picture when the particle has an entropic reason to be on one side
of the screen and the screen carries a temperature, it will
experience an effective macroscopic force due to the statistical
tendency to increase its entropy. This is described by
\begin{equation}\label{f}
F=T \frac{\triangle S}{\triangle x}.
\end{equation}
where $\triangle x$ is the displacement of the particle from the
holographic screen, while $T$ and $\triangle S$ are the temperature
and the entropy change on the screen, respectively. According to the
Unruh formula, the temperature in Eq. (\ref{f}) associated with the
holographic screen is
\begin{equation}\label{Ta}
k_BT= \frac{\hbar g}{2\pi c}.
\end{equation}
where $g$ represents the proper gravitational acceleration on the
screen which is produced by the matter distribution inside the
screen. Suppose we have a mass distribution $M$ which induces a
holographic screen $\mathcal {S}$ at some distance $R$ that has
encoded on it gravitational information. Suppose we have also a test
mass $m$ which is assumed to be very close to the holographic screen
as compared to its reduced Compton wavelength
$\lambda_m=\frac{\hbar}{mc}$. Assuming the holographic screen forms
a closed surface. The key statement is that we need to have a
temperature in order to have a force. One can think about the
boundary as a storage device for information. Assuming that the
holographic principle holds, the maximal storage space, or total
number of bits, is proportional to the area $A$. Let us denote the
number of used bits by $N$. It is natural to assume that this number
will be proportional to the area $A=4\pi R^2$. Thus we write
\begin{equation}\label{N}
A=NQ,
\end{equation}
where $Q$ is a fundamental constant which should be specified later.
Note that $N$ is the number of bits and thus for one unit change we
have $\triangle N=1$, hence from (\ref{N}) we find $\triangle A=Q$.
Motivated by Bekenstein's area law of black hole entropy, we assume
the entropy associated with the holographic screen obey the area
law, namely
\begin{equation}\label{S1}
S=\frac{A}{4\ell _p^2},
\end{equation}
where $\ell _p^2=G\hbar/c^3$ is the Planck length. Following
\cite{Ver} we also assume the energy on the holographic screen is
proportional to the mass distribution $M$ that would emerge in the
part of space enclosed by the screen
\begin{equation}\label{Ec}
 E=M c^2.
\end{equation}
According to statistical thermodynamics the equipartition law of
energy for free particle only valid for the situation in which the
kinetic energy of particle is much larger than the effective
interacting potential between them. Therefore, the equipartition law
of energy fails in the very low temperatures. It is found that Debye
model, which modified the equipartition law of energy, is in good
agreement with experimental results for most solid objects.
Following Verlinde's scenario, the laws of gravitation are no longer
fundamental, but rather emerge naturally as an entropic force. It
means that the gravity may have a statistical thermodynamics origin.
Thus, any modification of statistical mechanics should modify the
laws of gravity accordingly. Motivated by this point, we modify the
equipartition law of energy  as
\begin{equation}\label{E1}
 E=\frac{1}{2}Nk_B T\mathcal D(x),
 \end{equation}
where the  Debye function is defined by
\begin{equation}\label{D}
\mathcal D(x)\equiv\frac{3}{x^{3}}\int_ 0^x\frac{y^{3}}{e^{y}-1}dy.
\end{equation}
Here $x$ is related to the temperature $T$ as follows
\begin{equation}\label{x}
x\equiv\frac{T_D}{T}=\frac{\hbar \omega_D}{Tk_B},
\end{equation}
where $T_D$ is the Debye critical temperature and $\omega_D$ is the
Debye frequency. Combining Eqs. (\ref{N}), (\ref{Ec}) and
(\ref{E1}), we obtain the temperature of the holographic screen as
\begin{equation}\label{T2}
T=\frac{2Mc^{2}Q}{4\pi R^{2}k_B\mathcal D(x)}.
\end{equation}
Substituting Eqs. (\ref{T2}) and (\ref{S1}) in (\ref{f}), and using
relation $\triangle A=Q$, we get
\begin{equation}\label{F3}
F=-\frac{Mm}{R^2}\frac{1}{\mathcal D(x)}\left(\frac{Q^2c^3}{8 \pi
\ell _p^2k_B \hbar}\right)
\end{equation}
where we have taken $\triangle x=-\frac{\hbar}{mc}$ for one
fundamental unit change in the entropy and the entropy gradient
points radially from the outside of the surface to inside. In order
to derive the Newton's law of gravitation we must define $Q^2=8\pi
k_B \ell_p^4$. Finally we reach
\begin{equation}\label{F4}
F=-G\frac{Mm}{R^{2}}\frac{1}{\mathcal D(x)}.
\end{equation}
The corresponding gravitational acceleration is obtained as
\begin{equation}\label{g}
g=\frac{GM}{R^{2}}\frac{1}{\mathcal D(x)}.
\end{equation}
Using relation (\ref{Ta}) we can define the Debye acceleration $g_D$
which is related to the Debye temperature as
\begin{equation}
T_D=\frac{\hbar g_D}{2\pi k_B c}, \   \    \
x=\frac{T_D}{T}=\frac{g_D}{g}
\end{equation}
Eq. (\ref{F4}) is the Newton's law of gravitation resulting from
Debye entropic force. Let us study two different limits of Eq.
(\ref{F4}). In the strong gravitational field limit, i.e. at high
temperature, $T\gg T_D$ ($x\ll1$), the Debye function reduces to
\begin{equation}
\mathcal D(x)\approx 1.
\end{equation}
As a result in this limit, the usual Newtonian gravity is restored.
In the weak gravitational field limit, i.e. at very law temperature,
$T\ll T_D$ ($x\gg1$) we have
\begin{equation}\label{dx}
\mathcal D(x)=\frac{\pi^{4}}{5x^{3}}=\frac{\pi^{4}}{5}
\left(\frac{g}{g_D}\right)^{3}
\end{equation}
In this limit, the Newton's law is modified as
\begin{equation}\label{F6}
F=-\frac{5GMm}{R^{2}}\frac{g_D^{3}}{\pi ^{4}g^{3}},
\end{equation}
while the gravitational acceleration becomes
\begin{equation}
g=\left(\frac{5GMg_D^{3}}{\pi^{4}}\right)^{\frac{1}{4}}\frac{1}{\sqrt{R}}
\  \   \Rightarrow \  g\propto \frac{1}{\sqrt{R}}
\end{equation}
Therefore in this limit the gravitational field differs from
Newtonian gravity.
Let us then consider the cosmological implications of the presented
model. We assume the background spacetime is spatially homogeneous
and isotropic which is described by the line element
\begin{equation}
ds^2={h}_{\mu \nu}dx^{\mu} dx^{\nu}+R^2(d\theta^2+\sin^2\theta
d\phi^2),
\end{equation}
where $R=a(t)r$, $x^0=t, x^1=r$, the two dimensional metric $h_{\mu
\nu}$=diag $(-1, a^2/(1-kr^2))$. Here $k$ denotes the curvature of
space with $k = 0, 1, -1$ corresponding to flat, closed, and open
universes, respectively. The dynamical apparent horizon, a
marginally trapped surface with vanishing expansion, is determined
by the relation $h^{\mu \nu}\partial_{\mu}R\partial_{\nu}R=0$. A
simple calculation gives the apparent horizon radius for the
Friedmann-Robertson-Walker (FRW) universe
\begin{equation}
\label{radius}
 R=ar=\frac{1}{\sqrt{H^2+k/a^2}}.
\end{equation}
The matter source in the FRW universe is assumed as a perfect fluid
with stress-energy tensor
\begin{equation}\label{T}
T_{\mu\nu}=(\rho+p)u_{\mu}u_{\nu}+pg_{\mu\nu}.
\end{equation}
Now we are in a position to derive the dynamical equation for
Newtonian cosmology. Consider a compact spatial region $V$ with a
compact boundary $\mathcal S$, which is a sphere with physical
radius $R= a(t)r$. Note that here $r$ is a dimensionless quantity
which remains constant for any cosmological object partaking in free
cosmic expansion. Combining the second law of Newton for the test
particle $m$ near the surface, with gravitational force (\ref{F4})
we get
\begin{equation}\label{F5}
m\ddot{R}=m\ddot{a}r=-G\frac{Mm}{R^{2}} \frac{1}{\mathcal D(x)}.
\end{equation}
We also assume $\rho=M/V$ is the energy density of the matter inside
the the volume $V=\frac{4}{3} \pi a^3 r^3$. Thus, Eq. (\ref{F5}) can
be rewritten as
\begin{equation}\label{F6}
 \frac{\ddot{a}}{a}=-\frac{4\pi G}{3}\rho\frac{1}{\mathcal D(x)}
\end{equation}
This is the dynamical equation for Newtonian cosmology which is
valid in all range of the temperature. For strong gravitational
field ($\mathcal D(x)\simeq 1$) we reach the well-known formula
\begin{equation}\label{F7}
 \frac{\ddot{a}}{a}=-\frac{4\pi G}{3}\rho.
\end{equation}
Next we want to derive the Friedmann equations of FRW universe. For
this purpose we need to employ the concept of the active
gravitational mass $\mathcal M$ \cite{Pad3}, since this quintity
produces the acceleration in general relativity. From Eq. (\ref{F6})
with replacing $M$ with $\mathcal M$ we have
\begin{equation}\label{M1}
\mathcal M=-\frac{\ddot{a} a^{2}r^{3}}{G}\mathcal D(x)
\end{equation}
On the other side, the active gravitational mass is  defined as
\cite{Cai4}
\begin{equation}\label{Int}
\mathcal M =2
\int_V{dV\left(T_{\mu\nu}-\frac{1}{2}Tg_{\mu\nu}\right)u^{\mu}u^{\nu}}.
\end{equation}
A simple calculation gives
\begin{equation}\label{M2}
\mathcal M =(\rho+3p)\frac{4\pi}{3}a^3 r^3.
\end{equation}
Equating Eqs. (\ref{M1}) and (\ref{M2}) we obtain
\begin{equation}\label{addot}
\frac{\ddot{a}}{a}=-\frac{4\pi G}{3}(\rho+3p)\frac{1}{\mathcal D(x)}.
\end{equation}
This is the modified acceleration equation for the dynamical
evolution of the  FRW universe. Multiplying $\dot{a}a$ on both
sides of Eq. (\ref{addot}), and using the continuity equation
\begin{equation}\label{cont}
\dot{\rho}+3H(\rho+p)=0,
\end{equation}
after integrating we find
\begin{equation}\label{Fried2}
H^{2}+\frac{k}{a^{2}}=\frac{8\pi G}{3 a^2}\int \frac{d(\rho
a^{2})}{\mathcal D(x)}.
\end{equation}
This is the first Friedmann equation resulting from Debye entropic
force. Eqs (\ref{Fried2}) and (\ref{cont}) together with the
equation of state $p=w\rho$ govern the evolution of the universe. It
is important to note that since $x=x(T)$ and the temperature is also
a function of scale factor namely, $T=T(a)$, thus in general we
cannot integrate Eq. (\ref{Fried2}) and derive the simplified
result. When $ x\ll 1\Rightarrow\mathcal D(x)\approx 1$, the
well-known Friedmann equation in standard cosmology is recovered.
For $ x\gg 1$, using Eq. (\ref{dx}) we find
\begin{equation}\label{Fried3}
H^{2}+\frac{k}{a^{2}}=\frac{8\pi G}{3}\rho
\frac{5}{\pi^{4}}\left(\frac{g_D}{g}\right)^{3}.
\end{equation}
In order to derive the second Friedmann equation (\ref{Fried2}), we
have to combine the first Friedmann equation with continuity
equation (\ref{cont}). Let us put $k=0$ for simplicity, which has
been confirmed by recent observations. Differentiating Eq.
(\ref{Fried2}) we find
\begin{eqnarray}\label{Fried4}
 2H dH=\frac{8\pi G}{3}\left[-\frac{2}{a^{2}}\frac{da}{a}\int d(\rho a^{2})\frac{1}{\mathcal D(x)}
  +d\rho \frac{1}{\mathcal D(x)}+2\rho \frac{da}{a}\frac{1}{\mathcal D(x)}\right]
\end{eqnarray}
Multiplying $\frac{3}{2}$ on both sides of Eq. (\ref{Fried4}) and
dividing by $dt$, we find
\begin{eqnarray}\label{Fried5}
3H\dot{H}=-\frac{8\pi G}{a^{2}}\frac{\dot{a}}{a}\int d(\rho
a^{2})\frac{1}{\mathcal D(x)}+4\pi G\dot{\rho}\frac{1}{\mathcal
D(x)}+8\pi G\rho \frac{\dot{a}}{a}\frac{1}{\mathcal D(x)}.
\end{eqnarray}
Using the continuity equation (\ref{cont}), the above equation can
be written as
\begin{eqnarray}\label{Fried6}
-\left[\dot{H} \mathcal D(x)+\frac{8\pi G}{3a^{2}}\mathcal D(x)\int
d(\rho a^{2})\frac{1}{\mathcal D(x)}-\frac{8\pi
G}{3}\rho\right]=4\pi G(\rho+p).
\end{eqnarray}
For $ x\ll 1$  we have $D(x)\approx 1 $, and the well-known second
Friedmann equation of FRW universe in flat spacetime is recovered,
namely
\begin{eqnarray}\label{Fried7}
-\dot{H}=4\pi G(\rho+p).
\end{eqnarray}
It is worth noting that the Friedmann equation in Debye entropic
force scenario was first studied in \cite{Gao}. Let us stress the
difference between our derivation in this section and that of
\cite{Gao}. The author of \cite{Gao} has derived Friedmann
equations, following the method of \cite{FW}, by applying the
equipartition law of energy, $E =NT/2$, to the apparent horizon of
a FRW universe with the assumption that the apparent horizon has
the temperature $T=\hbar /(2\pi R)$, where $R$ is the apparent
horizon radius. Thus the total energy change of the system is
obtained as \cite{FW}
\begin{eqnarray}\label{dE}
dE=\frac{1}{2}NdT+\frac{1}{2}TdN=\frac{dR}{G},
\end{eqnarray}
during the infinitesimal time interval $dt$, where the apparent
horizon radius evolves from $R$ to $R+dR$. Indeed, the above
equation is just the first law of thermodynamics in the form
$dE=TdS$ on the apparent horizon,  where $T=\hbar /(2\pi R)$ and
$S=A/(4\hbar G)$ is the entropy of the system which assumed to
obey the area-law and $A=4\pi R^2$ is the apparent horizon area.
While in the present work we have not employed the first law of
thermodynamics for deriving the modified Friedmann equations.
Therefore, our result is independent of the definition of the
temperature in a dynamical spacetime.

\section{Entropic Corrected Friedmann Equation in Debye entropic gravity}\label{IV}
In this section we would like to consider the effects of the quantum
correction terms to the entropy expression, on the laws of gravity
in Debye model of entropic gravity. The result we will obtain are
valid in all range of temperature. The correction terms to the
entropy expression originate from the loop quantum gravity (LQG).
The quantum corrections provided to the entropy-area relationship
leads to the curvature correction in the Einstein-Hilbert action and
vice versa \cite{Zhu,Suj}. In the presence of quantum corrections
the entropy takes the following form \cite{Zhang}
\begin{equation}\label{Sc}
S=\frac{A}{4\ell _p^2}-\beta \ln {\frac{A}{4\ell _p^2}}+\gamma
\frac{\ell _p^2}{A}+\mathrm{const},
\end{equation}
where $\beta$ and $\gamma$ are dimensionless constants of order
unity. These corrections arise in the black hole entropy in LQG due
to thermal equilibrium fluctuations and quantum fluctuations
\cite{Rovelli}. We will show that these corrections modify the
Newton's law of gravitation as well as the Friedmann equations. First
of all we rewrite Eq. (\ref{Sc}) in the following form
\begin{equation}
\label{S2}
 S=\frac{A}{4\ell _p^2}+{s}(A),
\end{equation}
where $s(A)$ represents the correction terms in the entropy
expression. In this case the entropy change is obtained as
\begin{equation}
\label{S3}
 \triangle S=\frac{\partial S}{\partial A}\triangle
 A=\left(\frac{1}{4\ell _p^2}+\frac{\partial{s}(A)}{\partial A}\right)\triangle
 A.
\end{equation}
Substituting Eqs.  (\ref{T2}) and (\ref{S3}) in Eq. (\ref{f}) and
using relations $\triangle x=-\frac{\hbar}{mc}$ and $\triangle A=Q$,
one can easily find
\begin{equation}\label{Fc3}
F=-\frac{Mm}{R^2 \mathcal D(x)}\left(\frac{Q^2c^3}{2\pi k_B \hbar
}\right)\left[\frac{1}{4\ell _p^2}+\frac{\partial{s}}{\partial
A}\right]_{A=4\pi R^2}.
\end{equation}
If we define $ Q^{2} \equiv8\pi k_{B}\ell_p^{4}$,  as before, we
immediately derive the modified Newton's law of gravity in Debye
entropic gravity
\begin{equation}\label{Fc4}
 F=-G\frac{Mm}{R^{2}}\frac{1}{\mathcal D(x)}\left[1-\frac{\beta}{\pi}\frac{\ell_p^2}{R^2}
-\frac{\gamma}{4\pi^2}\frac{\ell_p^4}{R^4}\right].
\end{equation}
In the absence of correction terms ($\beta=\gamma=0$), the above
equation reduces to the result of the previous section. Let us study
two different limit of the above equation. In the strong
gravitational limit, i.e. at high temperature, $ T_D \ll T$
($\mathcal D(x)\approx 1)$ we have
\begin{equation}\label{Fc5}
 F=-G\frac{Mm}{R^{2}}\left[1-\frac{\beta}{\pi}\frac{\ell_p^2}{R^2}
-\frac{\gamma}{4\pi^2}\frac{\ell_p^4}{R^4}\right],
\end{equation}
which is exactly the result obtained in \cite{Sheykhi1}. When
$\beta=\gamma=0$, one recovers the well-known Newton's law. on the
other hand, at very law temperature $T_D\gg T$ we have $\mathcal
D(x) =\frac{\pi^{4}}{5x^{3}}$ and Eq. (\ref{Fc4}) reduces to
\begin{equation}\label{Fc6}
 F=-G\frac{Mm}{R^{2}}\frac{5}{\pi^{4}}\frac{g_D^{3}}{g^{3}}
 \left[1-\frac{\beta}{\pi}\frac{\ell_p^2}{R^2}
-\frac{\gamma}{4\pi^2}\frac{\ell_p^4}{R^4}\right].
\end{equation}
To derive Friedmann equation we follow the method of the previous
section. Combining the second law of Newton for the test particle
$m$ near the screen with gravitational force (\ref{Fc6}) we obtain
\begin{equation}\label{Fc7}
F=m\ddot{R}=m\ddot{a}r=-\frac{MmG}{a^{2}r^{2}}\frac{1}{\mathcal
D(x)}\left[1-\frac{\beta}{\pi}\frac{\ell_p^2}{R^2}
-\frac{\gamma}{4\pi^2}\frac{\ell_p^4}{R^4}\right]
\end{equation}
which from it we can derive the acceleration equation
\begin{equation}\label{ddotc2}
\frac{\ddot{a}}{a}=-\frac{4\pi G}{3}\rho \frac{1}{\mathcal
D(x)}\left[1-\frac{\beta}{\pi}\frac{\ell_p^2}{R^2}
-\frac{\gamma}{4\pi^2}\frac{\ell_p^4}{R^4}\right].
\end{equation}
With the entropic corrections terms, the active gravitational mass
$\mathcal M$ will be modified accordingly. The active gravitational
mass $\mathcal M$ in this case is obtained as
\begin{equation}\label{G2}
\mathcal M =-\frac{\ddot{a}a^{2}}{G}r^{3}\mathcal
D(x)\left[1-\frac{\beta}{\pi}\frac{\ell_p^2}{R^2}
-\frac{\gamma}{4\pi^2}\frac{\ell_p^4}{R^4}\right]^{-1}.
\end{equation}
Equating the above equation with Eq. (\ref{M2}) yields
\begin{equation}\label{D3}
\frac{\ddot{a}}{a}=-\frac{4\pi G}{3}(\rho+3p)\frac{1}{\mathcal
D(x)}\left[1-\frac{\beta}{\pi}\frac{\ell_p^2}{R^2}
-\frac{\gamma}{4\pi^2}\frac{\ell_p^4}{R^4}\right].
\end{equation}
Next we multiply the both sides of the above equation by $a\dot{a}$,
after using the continuity equation (\ref{cont}) and integrating we
find
\begin{eqnarray}\label{Friedc1}
 H^{2}+\frac{k}{a^{2}}&=&\frac{8\pi G}{3a^2}\left[\int \frac{d(\rho a^{2})}{\mathcal D(x)}
-\frac{\beta}{\pi}\frac{\ell_p^{2}}{r^{2}}
 \int \frac{d(\rho a^{2})}{a^{2}\mathcal D(x)} -\frac{\gamma}{4\pi ^{2}}\frac{\ell_p^{4}}{r^{4}}
 \int \frac{d(\rho a^{2})}{a^{4}\mathcal D(x)}\right].
\end{eqnarray}
Where $k$ is an integration constant. Unfortunately, the above
equation cannot be integrated in general for an arbitrary $\mathcal
D(x)$.  In the limiting case $D(x)\approx 1 $, the integrations can
be done following the method developed in \cite{Sheykhi1}. We find (
see \cite{Sheykhi1} for details)
\begin{eqnarray}\label{Friedc2}
 &&\left(H^2+\frac{k}{a^2}\right)+\frac{\beta \ell_p^2
(1+3\omega)}{3\pi(1+\omega)}\left(H^2+\frac{k}{a^2}\right)^{2}
+\frac{\gamma \ell_p^4 (1+3\omega)}{4\pi^2 (5+3\omega)}
\left(H^2+\frac{k}{a^2}\right)^{3}=\frac{8\pi G}{3}\rho.
 \end{eqnarray}
Again we see that in the absence of correction terms
$(\beta=0,\gamma=0)$ the well-known Friedmann equation is recovered.
For $x\gg 1$ ($\mathcal D(x)=\frac{\pi^{4}}{5x^{3}})$ Eq.
(\ref{Friedc1}) can be written
\begin{eqnarray}\label{h7}
H^{2}+\frac{k}{a^{2}}&=&\frac{8\pi G}{3}\left[\frac{1}{a^{2}}\int
d(\rho a^{2})\frac{5x^{3}}{\pi^{4}}\right. \nonumber
\\
&& \left. -\frac{\beta}{\pi}\frac{\ell_p^{2}}{a^{2}r^{2}}\int
\frac{d(\rho a^{2})}{a^{2}} \frac{5x^{3}}{\pi^{4}}
-\frac{\gamma}{4\pi ^{2}}\frac{\ell_p^{4}}{a^{2}r^{4}}\int
\frac{d(\rho a^{2})}{a^{4}}\frac{5x^{3}}{\pi^{4}}\right].
\end{eqnarray}
\section{conclusion}
Verlinde proposal on the entropic origin of the gravity is based
strongly on the assumption that the equipartition law of energy
holds on the holographic screen induced by the mass distribution
of the system. However, from the theory of statistical mechanics
we know that the equipartition law of energy does not hold in the
limit of very low temperature. By low temperature, we mean that
the temperature of the system is much smaller than Debye
temperature, i.e. $T\ll T_D$. It was demonstrated that the Debye
model is very successful in interpreting the physics at the very
low temperature. Since the discovery of black holes
thermodynamics, physicist have been thought that the gravitational
systems such as black hole and our universe can also be regarded
as a thermodynamical system. Hence, it is expected that the
equipartition law of energy for the gravitational system should be
modified in the limit of very low temperature (or very weak
gravitational field).

In this paper inspired by the Verlinde proposal and following the
Debye model of equipartition law of energy in statistical
thermodynamics, we modified the entropic gravity. First, we studied
the Debye entropic gravity and derived the modified Newton's law of
gravitation and the corresponding Friedmann equations which are
valid in all range of temperature. We found that the modified
entropic force returns to the Newton's law of gravitation while the
temperature of the holographic screen is much higher than the Debye
temperature. Then we extended our study to the case where there are
correction terms such as logarithmic correction in the entropy
expression. In this case we again reproduced the gravitational
equations for all range of temperature. Our study shows a deep
connection between Debye entropic gravity and modified Friedmann
equation. The microscopic statistical thermodynamical model of
spacetime may shed light on the origin of the Debye entropic gravity
and the microscopic origin for the Newton's law of gravity and also
Friedmann equations in cosmology.
\acknowledgments{This work has been supported financially by
Research Institute for Astronomy and Astrophysics of Maragha
(RIAAM), Iran.}

\end{document}